\documentclass[final,5p,times,twocolumn, sort&compress]{elsarticle}

\usepackage{graphicx,color}

\newcommand{\jwj}[1]{\textcolor{red}{#1}}

\graphicspath{{figs/}}

\journal{Carbon}

\begin{document}
\begin{frontmatter}

\title{Strain Engineering for Thermal Conductivity of Single-Walled Carbon Nanotube Forests}
\author[JWJ]{Jin-Wu Jiang\corref{cor}}
\ead{jwjiang5918@hotmail.com}
\cortext[cor]{Corresponding author}
\address[JWJ]{Shanghai Institute of Applied Mathematics and Mechanics, Shanghai Key Laboratory of Mechanics in Energy Engineering, Shanghai University, Shanghai 200072, People's Republic of China}

\begin{abstract}
We perform classical molecular dynamics simulations to investigate the mechanical compression effect on the thermal conductivity of the single-walled carbon nanotube (SWCNT) forest, in which SWCNTs are closely aligned and parallel with each other. We find that the thermal conductivity can be linearly enhanced by increasing compression before the buckling of SWCNT forests, but the thermal conductivity decreases quickly with further increasing compression after the forest is buckled. Our phonon mode analysis reveals that, before buckling, the smoothness of the inter-tube interface is maintained during compression, and the inter-tube van der Waals interaction is strengthened by the compression. Consequently, the twisting-like mode (good heat carrier) is well preserved and its group velocity is increased by increasing compression, resulting in the enhancement of the thermal conductivity. The buckling phenomenon changes the circular cross section of the SWCNT into ellipse, which causes effective roughness at the inter-tube interface for the twisting motion. \jwj{As a result, in ellipse SWCNTs, the radial breathing mode (poor heat carrier) becomes the most favorable motion instead of the twisting-like mode and the group velocity of the twisting-like mode drops considerably, both of which lead to the quick decrease of the thermal conductivity with further increasing compression after buckling.}
\end{abstract}


\end{frontmatter}


\section{Introduction}
Single-walled carbon nanotube (SWCNT) is a member from the carbon based nano-material family, which is characteristic for superior thermal conductivity.\cite{YuC2005nl,BalandinAA2008,Balandin2011nm} In many practical applications, the quasi-one-dimensional SWCNT is ensembled in a bulk form, such as the SWCNT mat (or buckypaper).\cite{EndoM2005nat,KakadeaBA2008carbon,SuppigerD2008carbon} Tubes in the SWCNT mat are randomly orientated, leading to the lose of SWCNT's advanced properties. For instance, the room temperature thermal conductivity in the SWCNT mat is typically in the range of $10^{-1}$ to 10 Wm$^{-1}$K$^{-1}$, which is several orders lower than the SWCNT.\cite{PrasherRS2009prl,VolkovAN2010prl,MahantaNK2010carbon,YangK2010jpcm,MemonMO2011carbon,AitkaliyevaA2013apl,AitkaliyevaA2013sr,WangJ2014apl}

Besides the randomly orientated SWCNT mat, there is a particular mat structure, with all tubes aligning parallel with each other. This structure is termed as SWCNT forest, or SWCNT crystalline. The vertically aligned SWCNT forest has promising practical applications such as supercapacitor electrodes for compact energy-storage devices, since it retains most mechanical, thermal, and electronic properties of the SWCNT.\cite{YuX2003ec,LauKKS2003nl,ZhangM2005sci,FutabaDN2006nm,VeeduVP2006nm} During the development of the SWCNT forest, the density used to be a bottleneck for its practical applications, because most advanced properties of the SWCNT forest will disappear if its density is too low.\cite{HuXJ2006jht,AkoshimaM2009jjap} Great progress has been achieved in recent years toward the production of dense SWCNT forest. Futaba et al. fabricated a high-density SWCNT forest by using the zipping effect of liquids to draw tubes together.\cite{FutabaDN2006nm} The density of the SWCNT forest is increased to $10^{12}$~{cm$^{-2}$} by Zhong et al. using the point-arc microwave plasma chemical vapor deposition.\cite{ZhongGF2006carbon} In 2010, a general catalytic chemical vapor deposition (CVD) design is developed to synthesize aligned forests of carbon nanotube with an area density above $10^{13}$~{cm$^{-2}$}, which can be further increased by using narrower tubes.\cite{EsconjaureguiS2010acsn} This value is already very close to the theoretical maximum density in a compact array.

During the application of dense SWCNT forests in the electronic device such as the supercapacitor electrodes, the thermal management becomes a key factor for its breakdown due to high power density. In this point of view, it is important to investigate the thermal transport capability of the SWCNT forest.\cite{XuJ2006,YanXH2006jap} The SWCNT forest inherits the good thermal transport properties in its axial direction from the SWCNT. Its thermal conductivity is around 1750-5800 Wm$^{-1}$K$^{-1}$ in the axial direction as estimated in the experiment.\cite{HoneJ1999prb} The 3-Omega measurements obtained the axial thermal conductivity of the SWCNT forest to be 83 Wm$^{-1}$K$^{-1}$ at 323~K, which should be higher with sample quality improvement.\cite{HuXJ2006jht} Zhang et al. demonstrated that the room temperature axial thermal conductivity in SWCNT forest is about 450~{Wm$^{-1}$K$^{-1}$}, and this value can be increased linearly by increasing forest density.\cite{ZhangL2012nl} Besides the axial thermal transport, the inter-tube thermal conductivity is also very helpful to pump thermal energy out off the electronic device. Although the SWCNT forest is able to deliver heat along its axial direction very fast, the inter-tube thermal conductivity in the SWCNT forest is extremely low due to the weak van der Waals inter-tube interaction.

\begin{figure}[tb]
  \begin{center}
    \scalebox{1.0}[1.0]{\includegraphics[width=8cm]{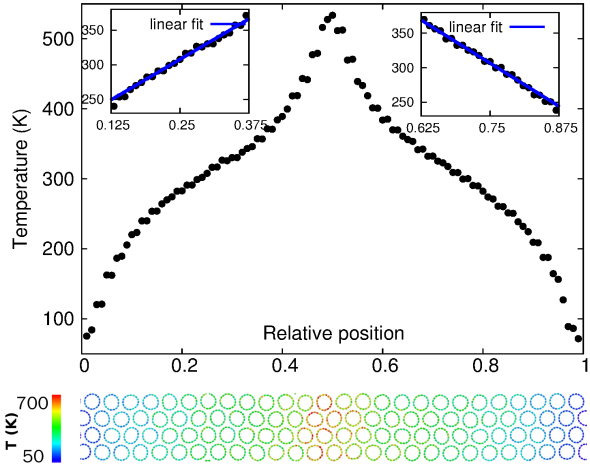}}
  \end{center}
  \caption{(Color online) Temperature profile for the SWCNT forest at room temperature. The current ratio $\alpha=0.01$ and the relaxation time of the heat bath is $\tau=0.04$~ps. The top left inset shows the linear fitting for the profile in $x\in[0.125, 0.375]$, giving a temperature gradient $\frac{dT}{dx}_{1}$. The right inset shows the linear fitting for the profile in $x\in[0.625, 0.875]$, giving a temperature gradient $\frac{dT}{dx}_{2}$. These two temperature gradients ($\frac{dT}{dx}_{1}$ and $\frac{dT}{dx}_{2}$) are averaged to give the temperature gradient in the computation of the thermal conductivity using the Fourier law. Bottom configuration shows the temperature distribution in the SWCNT forest. The color is with respective to the temperature of each carbon atom.}
  \label{fig_dTdx}
\end{figure}

To improve the situation, significant effort has been made to increase the inter-tube thermal conductivity.\cite{SinhaS2005jnr,MarconnetAM2011acsn,AitkaliyevaA2013apl,AitkaliyevaA2013sr} It was found that the thermal conductivity of the nanotube forest can be enhanced considerably by increasing its density.\cite{MarconnetAM2011acsn} Basically, the low inter-tube thermal conductivity is mainly because the van der Waals interaction between two SWCNTs is too weak. To enhance the inter-tube interaction, the ion irradiation technique was applied to introduce displacements between neighboring tubes, so the inter-tube thermal conductivity can be increased.\cite{AitkaliyevaA2013sr,AitkaliyevaA2013apl,WangJ2014apl}  Another straightforward solution to enhance the inter-tube van der Waals interaction is to compress the SWCNT forest using mechanical engineering approach. This forms the main scope of present work.

In this paper, we perform molecular dynamics simulations to examine the compression effect on the inter-tube thermal conductivity of the SWCNT forest.  We find that the thermal conductivity increases linearly with increasing compression before the SWCNT forest is buckled, because the inter-tube interaction is enhanced by compression. However, the thermal conductivity decreases quickly with further compression after the buckling of the SWCNT forest, where the cross section of the SWCNT changes from circular into ellipse. It is because the twisting (TW) mode is depressed in the buckled SWCNT forest, while the radial breath mode (RBM) becomes a favorable vibration mode. The RBM has a poor heat delivery capability, leading to the decrease of the thermal conductivity with increasing compression in the buckled SWCNT forest.

\begin{figure}[tb]
  \begin{center}
    \scalebox{1.1}[1.1]{\includegraphics[width=8cm]{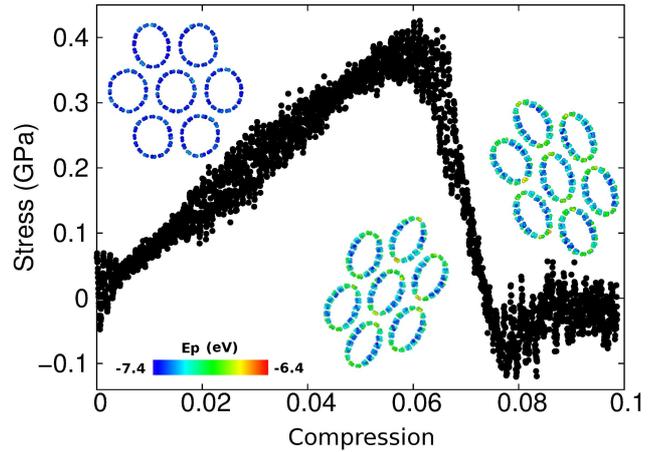}}
  \end{center}
  \caption{(Color online) Stress versus strain for SWCNT forest at room temperature. Buckling happens at compression with $\epsilon=-0.061$. Insets (left to right) show local configurations for SWCNT forest at compression with strain $\epsilon=$ -0.05, -0.07, and -0.08, respectively. Colors are with respective to the potential of each atom.}
  \label{fig_stress_strain}
\end{figure}

\begin{figure*}[tb]
  \begin{center}
    \scalebox{1}[1]{\includegraphics[width=\textwidth]{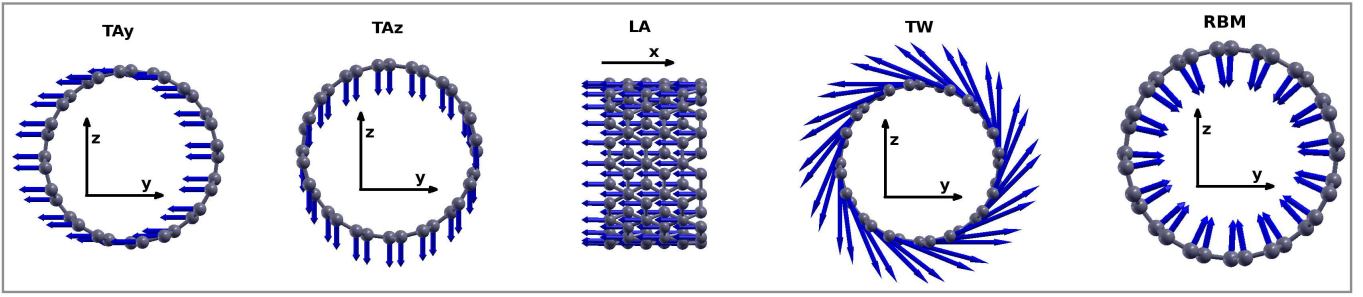}}
  \end{center}
  \caption{(Color online) Eigen vectors for five typical phonon modes in a SWCNT (8, 8). From left to right are the y-direction TA mode, z-direction TA mode, LA mode, TW mode, and the RBM. The arrow attached on each atom represents the vibrational displacement of each carbon atom in the corresponding phonon mode.}
  \label{fig_u}
\end{figure*}
\section{Structure and simulation details}
We study the the inter-tube thermal conductivity, which captures the ability of transporting heat energy from a SWCNT to its neighboring tubes in the forest. Our study focuses on the SWCNT forest that is constructed with armchair SWCNT (8, 8). All SWCNTs are closely packed and parallel to each other in the axial direction (see the bottom configuration in Fig.~\ref{fig_dTdx}). There are 26 SWCNTs aligned in the longitudinal direction and 4 SWCNTs in the lateral direction. The length of an individual SWCNT is 14.8~{\AA}, which forms the thickness of the SWCNT forest. \jwj{The size of the SWCNT forest is $393.3 \times 48.7 \times 14.8$~{\AA}. There are 21504 carbon atoms in the system. We have examined that the inter-tube thermal conductivity for the SWCNT forest will increase with increasing length for the system. This is similar as the the abnormal length dependence for the thermal conductivity of graphene, silicon nanowire, and other nano-materials.\cite{BalandinAA2008,LindsayL,Balandin2011nm,XuX2014nc}}

The thermal transport is simulated by the direct molecular dynamics simulation method implemented in the LAMMPS package, while the OVITO package was used for visualization in this section.\cite{ovito} The C-C interactions in the SWCNT forest are described by the Adaptive Intermolecular Reactive Empirical Bond Order (AIREBO) potential.\cite{StuartSJ2000jcp} The distance cut-off for the Lennard-Jones potential in the AIREBO is chosen as 1.02~{nm}. The standard Newton equations of motion are integrated in time using the velocity Verlet algorithm with a time step of 1~{fs}. Periodic boundary conditions are applied in all directions. 

The thermal current across the system is driven by frequently moving kinetic energy from the cold region to the hot region.\cite{IkeshojiT} This energy transfer is accomplished via scaling the velocity of atoms in the hot or cold temperature-controlled regions. The total thermal current pumped into the hot region (or pumped out from the cold region) is $J = \alpha E_{k}^{\rm hot}$. We have introduced a current parameter $\alpha$ to measure the energy amount to be aggregated per unit time. We note that the total current $J$ is the \textit{eflux} parameter in the \textit{fix heat} command in LAMMPS. The physical meaning of the parameter $\alpha$ is the ratio between the aggregated energy and the kinetic energy of each atom, i.e $\alpha$ measures the thermal current for each atom with respect to its kinetic energy.\cite{JiangJW2013sw} We have chosen a small parameter $\alpha=0.01$. A small $\alpha$ is very important for a successful simulation of inter-tube thermal transport, because the inter-tube thermal conductivity is so weak that atoms in cold regions will be frozen if their kinetic energies are pumped out too fast. The velocity scaling operation is performed every $\tau$ ps, which can be regarded as the relaxation time of this particular heat bath. We choose $\tau=0.04$~{ps} in all simulations.

After thermalization for a sufficiently long time, a temperature profile is established. Fig.~\ref{fig_dTdx} shows the temperature profile for the SWCNT forest at room temperature. The two insets show that the left middle region with relative axial position in $[0.125, 0.375]$ and the right middle region within $[0.625, 0.875]$ are linearly fitted to extract two temperature gradients. These two temperature gradients are averaged to give the final temperature gradient $\frac{dT}{dx}$. The thermal conductivity is obtained through the Fourier law,
\begin{eqnarray}
\kappa = \frac{1}{2} \times \frac{J}{|dT/dx|},
\label{eq_fourier}
\end{eqnarray}
where a factor of $\frac{1}{2}$ comes from the fact that heat current flows in two opposite directions. The difference between the left and right temperature gradients is used to estimate the numerical error in the thermal conductivity.

\section{Results and discussions}
Fig.~\ref{fig_stress_strain} shows the stress-strain relation for the SWCNT forest during mechanical compression at room temperature. The structure is compressed by uniformly deforming the structure in longitudinal direction with a strain rate of $\dot{\epsilon}=10^{8}$~s$^{-1}$, which is a typical value in MD simulations as shown in previous works.~\cite{JiangJW2012jmps} The structure is allowed to fully relax in transverse directions. Insets illustrate representative local configurations of the SWCNT forest, which is compressed at strains of $\epsilon=$ -0.05, -0.07, and -0.08, respectively. The buckling of SWCNTs in the forest starts at $\epsilon=-0.061$. After buckling, the cross section of the SWCNT turns into ellipse and the stress concentration becomes obvious at the two narrow corners of the ellipse, which is indicated by more bright color.

\begin{figure}[tb]
  \begin{center}
    \scalebox{1.0}[1.0]{\includegraphics[width=8cm]{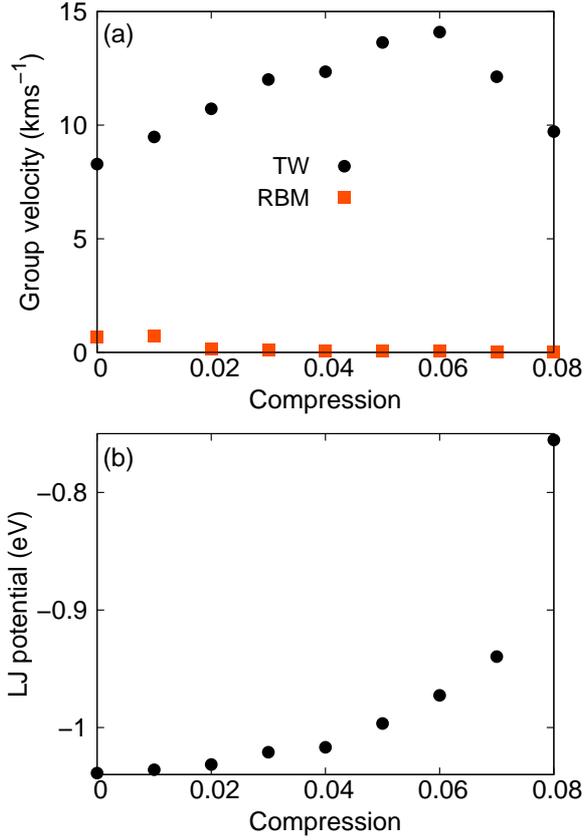}}
  \end{center}
  \caption{(Color online) Compression effect on (a) the phonon group velocities for the TW mode and the RBM, and (b) the inter-tube van der Waals interaction.}
  \label{fig_velocity}
\end{figure}

Phonon modes are the solely thermal energy carrier in the lattice thermal transport simulated in present work, so it is necessary to illustrate some associated phonon modes. Fig.~\ref{fig_u} shows the eigen vector of five typical phonon modes in the SWCNT (8, 8). The blue arrow on top of each atom represents the vibrational displacement of each carbon atom in the corresponding phonon mode. The first four panels represent the four acoustic modes, i.e, the y-direction transverse acoustic (TAy), the z-direction transverse acoustic (TAz), the longitudinal acoustic (LA), and the TW mode. The last panel shows the RBM. \jwj{The frequency of the RBM is inversely proportional to the tube diameter and its group velocity is almost zero as shown in Fig.~\ref{fig_velocity}~(a),\cite{PopovVN1999prb,PopovVN2000prb,JiangJW2006,JiangJW2008} so this mode possesses poor capability in delivering thermal energy.}

\begin{figure}[tb]
  \begin{center}
    \scalebox{1.0}[1.0]{\includegraphics[width=8cm]{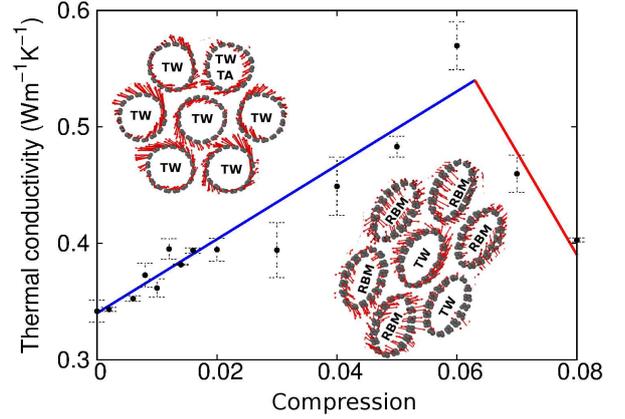}}
  \end{center}
  \caption{(Color online) Compression effect on the thermal conductivity of SWCNT forest at room temperature. Left top inset shows an instant local configuration for the SWCNT forest at compression $\epsilon=-0.05$. Red arrows attached on each atom denote the velocity of carbon atoms. The texts TW and TA imply that SWCNTs are involving in movements similar as the TW mode or the TA mode. Right bottom inset shows an instant local configuration for the SWCNT forest at compression $\epsilon=-0.07$, where the cross section of the SWCNT becomes ellipse. The movements of SWCNTs are similar as the TW mode or the RBM. Lines are guide to the eyes.}
  \label{fig_thermal_conductivity}
\end{figure}

To examine the compression effect on the thermal conductivity of the SWCNT forest, we present the compression dependence of the thermal conductivity in Fig.~\ref{fig_thermal_conductivity}. The SWCNT forest is compressed to a given strain value prior to the heat transport simulation. The thermal conductivity is then simulated with the non-equilibrium molecular dynamics method as outlined in previous section. We pay special attention to the modulation of the thermal conductivity by the buckling phenomenon at $\epsilon=-0.061$. It shows that the mechanical compression can enhance the thermal conductivity linearly before the occurrence of buckling phenomenon at $\epsilon=-0.061$, which is different from the robust thermal conductivity in the graphene under compression.\cite{WeiN} \jwj{This is due to the enhancement of the inter-tube van der Waals interaction. Fig.~\ref{fig_velocity}~(b) shows that the inter-tube van der Waals interaction increases with compression, resulting from shorter inter-tube spaces in the compressed SWCNT forest.}

 Left top inset illustrates an instant local configuration of the SWCNT forest at compression with $\epsilon=-0.05$ before buckling. The red arrow on top of each atom represents the velocity of the atom. After comparing this configuration with the eigen vector of the TW mode in Fig.~\ref{fig_u}, one finds that the vibrations of most SWCNTs are similar as the eigen vector of the TW phonon mode; i.e, the TW mode plays an important role for heat transport in the forest. It also implies that the TW-like movement has lower energy threshold, as the interface between two circular SWCNTs is very smooth. As a result, the TW-like movement is easy to be excited. \jwj{Fig.~\ref{fig_velocity}~(a) shows that the group velocity of the TW mode is at least two orders larger than the group velocity of the RBM, so the TW mode is a good heat carrier. The mechanical compression enhances the inter-tube interaction, and it does not modify the circular shape of the cross section for SWCNTs in the forest. As a result, the TW-like movement is still easy to be excited. Fig.~\ref{fig_velocity}~(a) shows that the phonon group velocity of the TW mode increases with increasing compression, which leads to the enhancement of the thermal conductivity with increasing compression.}

On contrary, Fig.~\ref{fig_thermal_conductivity} also shows that, for larger compression with $|\epsilon|>0.061$, the thermal conductivity decreases quickly with increasing compression. Right bottom inset illustrates an instant local configuration of the SWCNT forest at compression with $\epsilon=-0.07$ after buckling phenomenon. As compared with the local configuration before buckling, a distinct feature in the buckled structure is that most SWCNTs are involved in a motion that is similar as the eigen vector of the RBM shown in Fig.~\ref{fig_u}. The TW-like movement has been strongly depressed, because SWCNTs with ellipse cross section are difficult to twist, as a result of the effectively rough interface between two neighboring ellipse SWCNTs. \jwj{The RBM-like movement becomes a favorable vibration mode for the SWCNT with ellipse cross section in the forest. It is well known that the RBM of the SWCNT has higher frequency (inversely proportional to the tube diameter) than the TW acoustic mode (almost zero), and the RBM has very low group velocity as shown in Fig.~\ref{fig_velocity}~(a). Hence, the RBM-like movement is actually not benefit for transporting thermal energy. By increasing compression, more SWCNTs will participate the RBM-like motion, resulting in the decrease of the thermal conductivity of the SWCNT forest after buckling phenomenon. From Fig.~\ref{fig_velocity}~(a), we can find another mechanism for the decrease of the thermal conductivity of the SWCNT forest after buckling. It shows that the group velocity of the TW mode drops with compression after buckling, which also leads to the decrease of the thermal conductivity with compression after buckling.}

\jwj{In practice, the Joule heating is inevitable in the supercapacitor electrodes for compact energy-storage devices, due to the high power density. The SWCNT forest will be burnt down by such localized heating, if the thermal energy is not pumped out efficiently. Our study shows that inter-tube thermal conductivity of the SWCNT forest can be enhanced by applying mechanical compression before the buckling of the SWCNT. That is a smaller mechanical compression is helpful for the protecting the SWCNT forest supercapacitor from being burnt down, by dissipating the thermal energy from a SWCNT into its neighboring SWCNTs. However, a large mechanical compression is harmful for the heat delivery in the SWCNT forest.}

\section{Conclusion}
We have performed classical molecular dynamics simulations to study the thermal conductivity of the compressed SWCNT forest. The mechanical compression can enhance the thermal conductivity before the buckling of the SWCNT in the forest, but it weakens the thermal conductivity after the buckling happens. The origin for this thermal conductivity manipulation is discussed based on the interplay between the TW mode (good heat carrier) and the RBM (poor heat carrier).

\section*{Acknowledgements} The work is supported by the Recruitment Program of Global Youth Experts of China and the start-up funding from Shanghai University. The author thanks Ai-Jun Li in SHU for valuable comments.


\end{document}